**Data Analysis and Modeling for Transitioning Between Laboratory Methods for Detecting SARS-CoV-2 in Wastewater**

**Authors:** Maria M. Warns, Leah Mrowiec, Christopher Owen, Adam Horton, Chi-Yu Lin, Modou Lamin Jarju, Niall M. Mangan, Aaron Packman, Katelyn Plaisier Leisman, Abhilasha Shrestha, Rachel Poretsky

**Abstract**

Wastewater surveillance has proven to be a useful tool to monitor pathogens such as SARS-CoV-2 as it is a nonintrusive way to survey the potential disease burden of the whole population contributing to a sewershed. With the expansion of this field since the beginning of the COVID-19 pandemic, laboratory methods to process wastewater and quantify pathogen nucleic acid levels have improved as technologies changed, efforts expanded in size and scope, and supply chain issues were resolved. Maintaining data continuity is crucial for labs undergoing method transitions to accurately assess infectious disease levels over time and compare measured RNA concentrations to public health data. Despite the dynamic nature of laboratory methods and the necessity to ensure uninterrupted data, to our knowledge there has not been a study that unites two datasets from different lab methods for pathogen quantification from environmental samples. Here, we present a novel, data science-based way to resolve this research bottleneck by describing a lab transition from SARS-CoV-2 RNA quantification using a low-throughput, manual filtration-based wastewater concentration and RNA extraction followed by qPCR to a high-throughput, automated magnetic bead-based concentration and extraction followed by dPCR. During the two-month transition period, wastewater samples from across the Chicago metropolitan area were processed with both methods in parallel. We evaluated a variety of regression models to relate the RNA measurements from both methods and found that a log-log model was most appropriate after removing outliers and discrepancy points to improve model performance. We also evaluated the consequences of assigning values to samples that were below the detection limit. Our study demonstrates that data continuity can be maintained throughout a transition of laboratory methods if there is a sufficient period of overlap between the methods for an appropriate model to be constructed to relate the datasets.

**Introduction**

The use of wastewater surveillance has expanded over the past several decades to become an approach for nonintrusive monitoring of pathogens such as SARS-CoV-2. Work in this field began in the 1940s with the testing of sewage in United States cities for poliovirus (Metcalf et al. 1995), and has since tracked both pharmaceutical and illicit drug use (Duan et al. 2022, Gushgari et al. 2019) and viruses including influenza, mpox, and norovirus (Berchenko et al. 2017, Wolfe et al. 2022, Wolfe et al. 2023, Kazama et al. 2016). The COVID-19 pandemic heightened interest in wastewater surveillance due to the discovery that both asymptomatic and symptomatic individuals shed SARS-CoV-2 RNA in their stool, urine and other bodily fluids (Li et al. 2020, Chen et al. 2020).

SARS-CoV-2 RNA can be quantified in wastewater by polymerase chain reaction (PCR) methods such as digital PCR (dPCR), droplet digital PCR (ddPCR), and quantitative PCR (qPCR) to provide an additional estimate of prevalence that can supplement traditional public health indicators like hospital data and public testing program case numbers (Leisman et al. 2024, Bibby et al. 2021, Olesen et al. 2021). The shifting use and reporting of at-home testing, the end of the pandemic emergency, and decreased hospitalizations due to rising immunity have decreased the utility of traditional epidemiological data. Wastewater testing is less affected by these challenges since it surveys the whole population contributing to the sewershed and is not susceptible to changes in at-home and clinical testing. Today, wastewater is often the only measured prevalence indicator for SARS-CoV-2 and thus it is valuable to have an approach to maintain data continuity without relying on other public health measures of COVID-19 prevalence. Our approach is applicable to other nucleic acid biomarkers as well.

As wastewater surveillance programs have expanded and new technologies have become available, analytical approaches have improved. Maintaining data continuity is critical during methodological transitions in order to have accurate longitudinal datasets. To our knowledge, there has not been a documented way to integrate data generated by different lab methods for SARS-CoV-2 RNA quantification into a single dataset or avoid gaps in data during these transitions. However, there have been numerous published examples of comparisons between laboratory methods for SARS-CoV-2 detection in wastewater (Pérez-Cataluña et al. 2021, Chik et al. 2021, Zheng et al. 2022, Kevill et al. 2022). Here, we describe a methodological transition with a two month overlap period in which we shifted our approach from a low-throughput, manual process involving concentration of wastewater by filtration, manual RNA extraction, and qPCR to a high-throughput, automated process involving magnetic bead-based concentration and extraction followed by dPCR.

A transition from our manual low-throughput methods (LT) to automated high-throughput methods (HT) was necessary as we expanded our surveillance efforts from 30 samples per week from 15 Chicago locations to almost 200 samples per week from 80 locations across Illinois. There are notable differences between the two methods. Our LT method requires replicates, a standard curve, a greater sample volume (25 mL), and is limited by a low-throughput filtration step that processes only 6-8 samples at a time and can occur over a period of several hours depending on the amount of particulate matter. Because the LT method uses larger volumes, it may be more effective for low-abundance targets. Our HT methods are more cost effective and efficient, as they give us the ability to process a greater number of samples simultaneously, do not require filtration, and require a smaller sample volume (10 mL). Additionally, the new method improved our ability to carry out whole-genome sequencing by providing higher quality RNA that produced high coverage of SARS-CoV-2 genomes (Feng et al. 2023, von Wintzingerode et al.1997, Cullen et al. 2022). Furthermore, the Centers for Disease Control (CDC) has emphasized the utility of the HT method with magnetic beads and has endorsed the use of dPCR for the National Wastewater Surveillance System (Adams et al. 2024, Centers for Disease Control and Prevention 2023b). Through this transition study we obtained a large overlapping dataset and generated a regression model relating measurements from the LT and HT methods, preserving continuity in the surveillance time-series.

### Wastewater Sample Collection

Raw 1-liter wastewater samples were collected from 17 locations across the Chicagoland area including sewers, facilities, and wastewater reclamation plants (WRPs). Four locations were wastewater reclamation plants that provided time-proportional 24-hour flow-weighted composite samples and the remainder of the locations provided grab samples from facilities and sewers where composite sampling was not possible. In total, 315 samples were collected over a two-month period from November-December 2021. Samples were transported to the labs on ice. Upon arrival, samples were aliquoted into 50 mL falcon tubes for processing and analysis with the LT and HT approaches.

### Wastewater Concentration and Nucleic Acid Extraction

The LT method is described in Owen et al. (2022). In short, 25 mL of wastewater were spiked with 40 µL of lyophilized Bovine Coronavirus that had first been resuspended in 1mL TE and then diluted 1:100 in deionized (DI) water (BCoV; Zoetis Services, LLC) as a recovery control. Samples were then concentrated using $MgCl_2$ (25 nM) precipitation followed by filtration onto a 0.8-µm pore size mixed cellulose ester (MCE) membrane filter. RNA was then extracted from the sample filters using bead-beating and the QIAmp Viral RNA MiniPrep kit (Qiagen) following the manufacturer's protocol with slight modifications, using the filter membranes instead of the 140 µL liquid sample and including carrier RNA.

The HT method was adapted from Karthikeyan et al. (2021). Briefly, 10 mL of wastewater was spiked with 80 µL of BCoV solution and then concentrated on the KingFisher Apex Purification System (Thermo Fisher Scientific) using magnetic Nanotrap Microbiome A Particles (Ceres). Total nucleic acids were then extracted on the KingFisher Apex using the MagMax Wastewater Ultra Nucleic Acid Isolation Kit (Thermo Fisher Scientific).

### SARS-CoV-2 Quantification

SARS-CoV-2 RNA was quantified at the N1 and N2 loci of the nucleocapsid (N) gene using primers and probes published by the United States Centers for Disease Control (Lu et al. 2020). BCoV was quantified via the transmembrane (M) gene with primers and probes from Decaro et al. 2008. The LT method used a one-step reverse-transcription quantitative polymerase chain reaction (RT-qPCR) as described in Owen et al. (2022). These same loci were quantified with the HT method using digital PCR with 10 µL of QIAcuity One-Step Viral RT-PCR Master Mix 4X (Qiagen), 0.4 µL of QuantiNova Multiplex Reverse Transcription Mix (QIAGEN), 2 µL of 20X N1-N2-BCoV Assay Solution for QIAcuity (GT Molecular), 7.6 µL of molecular grade water and 5 µL extracted RNA diluted 1:4 with molecular grade water for a total volume of 20 µL per sample. dPCR was done on a QIAcuity instrument in 26k 24-well nanoplates (Qiagen). For both methods, N1 and N2 results were reported as gene copies per liter (GC/L) of starting wastewater and BCoV was reported as a percent recovery. Percent recovery was calculated as the ratio of BCoV gene copies in the sample to the BCoV gene copies in the reference control.

**Data Cleaning and Analysis**

We performed data cleaning to ensure our data set was as accurate as possible. A total of 307 samples were analyzed with both approaches, and 16 of those samples were processed in duplicate with the HT method. Duplicate samples were listed as two separate lines in our data set. These 16 samples have different N1 and BCoV recovery rate values for the HT method but the same values for the LT method. We removed 22 HT samples where the BCoV recovery rate was 0, which typically implies an irregularity in sample processing. The majority of these (21) occurred during the first week of using the HT approach. Following CDC guidelines, we assigned non-detect data points a value of half the limit of detection, which is 33.5 GC/L for the LT method and 2040 GC/L for the HT method (Centers for Disease Control and Prevention 2023a). Our final data set had 602 unique observations, 301 from each method (Figure 1A).

For both the LT and HT methods, the concentrations measured during the study period are within the range of the samples measured before the study period (Figure 1B; historical dataset) or after the study period (Figure 1C; reference dataset), respectively. However, they are not a statistically representative sample due to the difference in disease dynamics that were measured in varying phases of the pandemic. Even so, the study data is sufficient to develop a model relating the two methods.

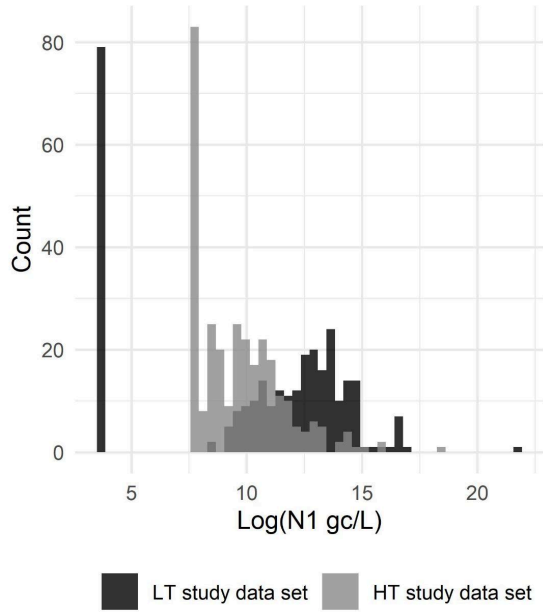

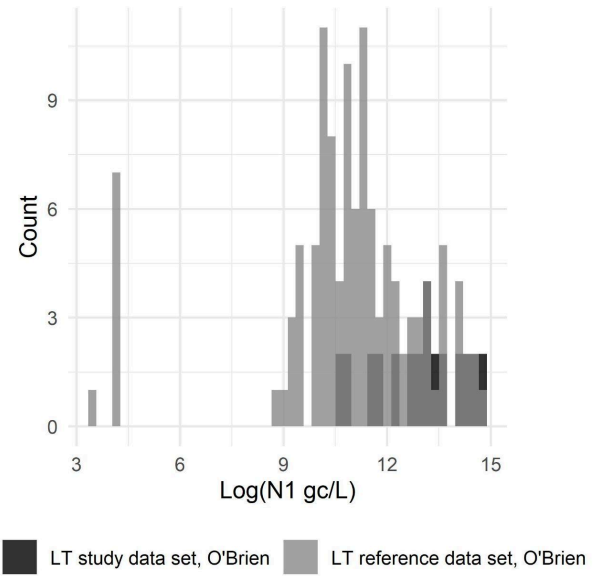

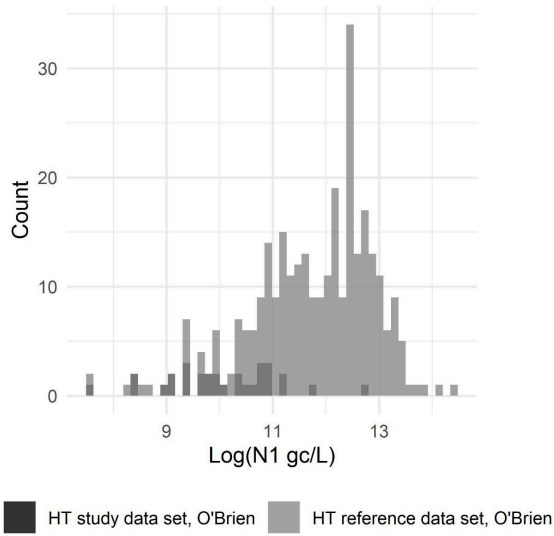

**Figure 1. A:** Histograms of the Log of N1 GC/L LT (dark grey) and HT (light grey) methods with overlap (medium grey) from all sampling locations. Non-detect values as half-LOD (two first columns of the distributions). **B:** Histogram of LT N1 GC/L from the O'Brien Wastewater Reclamation Plant from November 2020 to December 2021 (LT reference dataset, light) and from the period of methods transition, November 2021 to December 2021 (LT study dataset, dark). **C:** Histogram of HT N1 GC/L from O'Brien

Wastewater Reclamation Plant from February 2022 to September 2023 (HT reference dataset, light) and from the period of methods transition, November 2021 to December 2021 (HT study dataset, dark).

**Results: Model Selection**

We tested a series of regression models to relate the measurements from the two lab methods. All models were evaluated using the lm() function in R, with HT N1 GC/L as the predictor and LT N1 GC/L as the response variable (R Core Team, 2021). We found that an outlier made linear, quadratic, and exponential models appear to fit the data very well when only looking at metrics of model performance, but when that outlier was removed, all had adjusted R-squared values below 0.4. The Normal Q-Q plot for these models showed a lack of normality in the data and implied an inappropriate model structure. A log-log model, log(LT) ~ log(HT), applied to all 301 values had an adjusted R-squared of 0.50. We then calculated Cook's distances, a measure for outlier detection, for each sample, systematically removed points with the highest distances, and then recalculated the model (R Core Team, 2021). Points where one lab detected RNA and the other had a "non-detect" had the highest Cook's distance. "Non-detects" included any measurements below the nominal limit of detection (LOD) based on statistical analysis of calibration data. When we removed all 42 pairs with one non-detect and one detection, leaving N = 259 samples, the resulting model had an adjusted R-squared value of 0.65 (Equation 1; Figure 2B.):

$$(LT) = e^{-6.30 \pm 1.82} (HT)^{1.78 \pm 0.18} \tag{1}$$

**Non-detects in both Labs**

The parameters of this model and the resulting adjusted R-squared value depends strongly on what concentration is assigned to non-detects, given the large number (60) of non-detects in the final data set of 259. The CDC recommends using half the LOD (Centers for Disease Control and Prevention 2023a). The LT method has an LOD of 67 GC/L, while the HT method has an LOD of 4080 GC/L; for which the CDC recommendation yields assigned non-detect concentrations of 33.5 GC/L and 2040 GC/L, respectively. These values produced the regression model relating LT and HT data (Equation 1). However, the true concentration of each non-detect and the true distribution of concentrations between 0 GC/L and the LOD are unknown. Thus, any simple rule assigning a concentration to a non-detect overvalues the certainty of those points and potentially biases our model fit.

To evaluate the impact on the statistical correction model, we assumed below-detection-level concentrations were drawn from different beta distributions with varying mean and skew, set by $(\alpha, \beta)$ parameters using Monte Carlo methods (Figure 2. A.). We assume that the distribution shape is the same for each lab.

For each beta distribution shape, we randomly selected one ND concentration between 0 to the LOD for each measurement method, producing an ND-ND pair. We selected 60 such ND-ND pairs and refit the log-log model to the dataset with these pairs included (i.e., replacing the CDC-suggested values with the ND-ND pairs). We then repeated this random selection and fit 1000 times for each beta distribution, and calculated the mean and adjusted R-squared of the resulting best-fit model parameters (Figure 2. A.).

A

| Beta Distribution Parameters $(\alpha, \beta)$ | Slope | Intercept | Adjusted R-squared |
|---|---|---|---|
| (1, 3) | 1.69 | -5.88 | 0.72 |
| (2, 3) | 1.76 | -6.64 | 0.68 |
| (1, 1) | 1.70 | -6.05 | 0.64 |
| (3, 2) | 1.75 | -6.63 | 0.63 |
| (3, 1) | 1.72 | -6.46 | 0.60 |

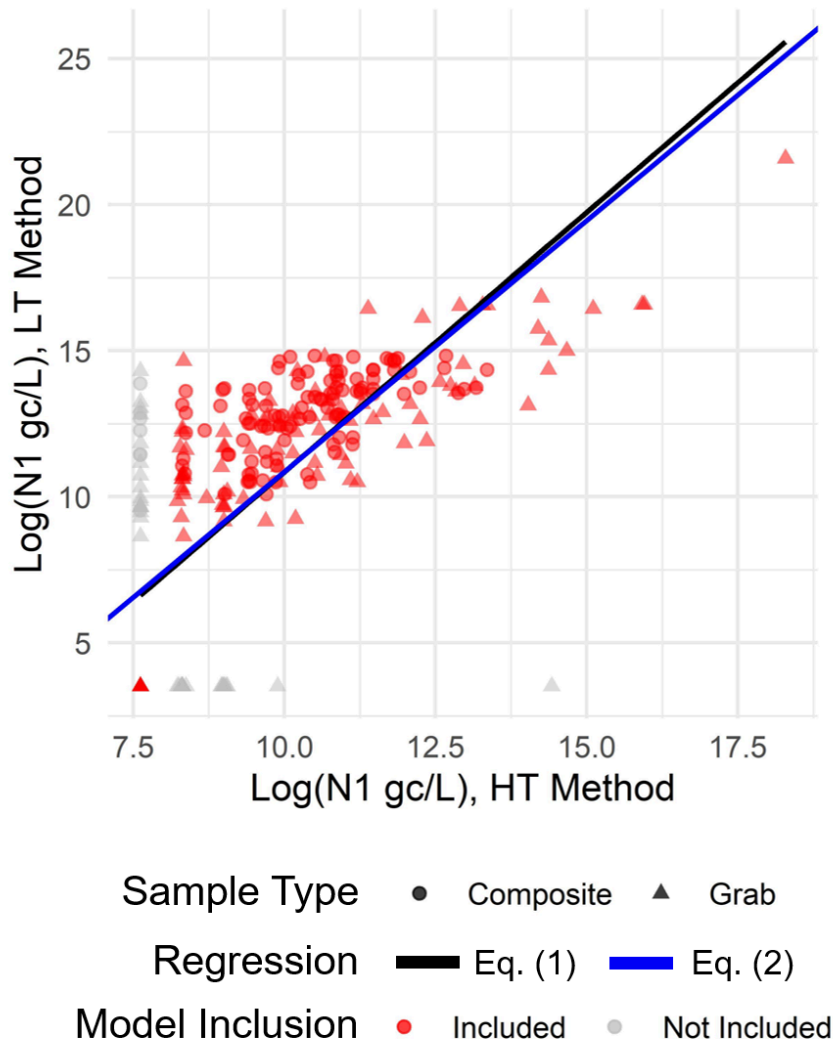

Figure 2. A: Averaged parameter values and adjusted R-squared for $(LT) = e^{intercept}(HT)^{slope}$. B: Regression fit for the study for both equations, comparing Eq. 1 and Eq. 2.

By varying the assumed distribution of the non-detects, we capture the high uncertainty below the limit of detection while still including the knowledge that a non-detect was measured in the model. We take the average of the slope and intercept across the beta distributions to form a final model incorporating this uncertainty (Equation 2). These parameter values are within the 95% confidence interval of those in Equation (1).

$$(LT) = e^{-6.33}(HT)^{1.72} \tag{2}$$

**Meaning and Utility**

Our study provides a novel approach to preserve data continuity in wastewater surveillance time-series when undergoing common transitions between laboratory methods as programs expand and new methods become available (Adams et al. 2024, Leonard 2023). Data consistency enables trend comparison over time and continuity of the relationship between RNA concentration ranges to public health data, regardless of sample processing or quantification techniques. Without accounting for transition effects, false trends can appear in measured SARS-CoV-2 RNA due to changes in methods and not in actual COVID-19 prevalence. By maintaining a period of overlap between lab techniques and creating a quantitative model to relate pre- and post-change datasets, researchers can maintain effective and continuous wastewater surveillance regardless of ever-changing laboratory methods.

**Discrepancies in SARS-CoV-2 Detection Results**

There were 42 instances where one method had a quantifiable detection and the other method had a non-detect. Of these instances, 19 were non-detects with the LT method and 23 were non-detects with the HT method. All cases where the LT had a non-detect and HT had a detection were grab samples, 89% (17) of which came from divisions in the Cook County Jail and 10% (2) from O'Hare International Airport. Among the samples where the HT method yielded a non-detect and the LT method had a detection, 52% (12) of the samples were grab samples from the Cook County Jail, 17% (4) were grab samples from O'Hare airport, and 30% (7) were composite samples from wastewater reclamation plants. The ubiquity of grab samples as outliers from both labs suggests that grab samples yield high variation regardless of the method used. This indicates a need to better understand the inherent variability of grab sampling. When a project undergoes a transition, it is critical that all collection methods are analyzed in an overlap period.

We also investigated the BCoV recovery rate as a possible indicator for samples with discrepancy between lab results. We found that 47% (9) of the samples detected with the HT but not the LT method had BCoV recovery rates over 100% for the HT lab results, implying potential irregularity in either the wastewater samples or the sample processing. Similarly, 56% (13) of the samples detected by the LT but not the HT method had a BCoV recovery rate over 100% for the LT lab results. However, for all samples with detection discrepancy, Welch's t-test did not indicate that their BCoV recovery rate was significantly different from that of the study set for the model in Equation (1). Thus a BCoV measurement of over 100% was not a reliable indicator for discrepant samples.

**Interpretations and Recommendations**

It is important for labs undergoing a method transition to test as many samples as possible with both methods and using samples that represent a wide range of gene concentrations. Due to the substantial impact that the value assigned to non-detect points has on model parameters and, consequently, measures of model fit, any estimates on a minimum number of data points required for such a study are only based on the number of detections. It is not possible to forecast perfectly when there will be a sufficiently broad range of nucleic acid concentrations in wastewater samples, so we encourage labs to be prepared to extend the comparison study period when too many samples are non-detects. Inherent randomness in wastewater measurements means it is impossible to predict exactly how long one should conduct a similar study when changing laboratory methods. It is possible that if a data set was small or poorly sampled the underlying distribution, the data could imply that a different model functional form was more parsimonious.

Recognizing that with program constraints like budget limitations, it is not always possible to have a long overlap data set, we investigated how many samples were necessary to develop a robust model with our data. We conducted a bootstrap analysis of our data set, beginning with a data set of 199, which included only the points from the study for which both labs had a detection. Varying the number of data points included in the log-log regression from 2 to the full data set size, we uniformly randomly selected the data points from our data set with replacement, then refit the log-log model. We repeated this process 1000 times for each data set size, from 2 to 199. We calculated the mean and 95% confidence interval for the two fit parameters and for each option of data set size. The change in the width of the confidence interval is largest before approximately 25 data points. With at least 51 points, the change in the 95% confidence bounds is consistently less than 0.01. We advise groups undergoing a similar transition of laboratory methods to fit a regression model as each data point is added, excluding non-detects. The study should continue until there are a sufficient number and range of points such that the model parameters and their confidence intervals reach a plateau.

Future research is needed to more precisely quantify the variability of wastewater samples as well as the variability generated by field sampling methods and lab techniques. It would be useful to further investigate which step(s) in the methods the variability is coming from or if it is due to the inherent heterogeneity and variability of wastewater. The data analysis and modeling in this study is limited to SARS-CoV-2 data, but the overall approach could be applied to any target tested under changing lab methods, and such a study could elucidate how different methods obtain different concentrations from the same sample across targets.

**Conclusions**

- ❖ Data continuity during a transition of lab methods can be maintained and research bottlenecks can be avoided if labs carefully design a transition period where samples are analyzed under both methods for as many samples as possible
- ❖ Analysis of outliers and removal of such inconsistent points improves model performance
- ❖ Given the highly variable nature of wastewater, longer periods of overlap with both quantification techniques and a larger number of samples analyzed enables better identification of outlier points and a wider domain for a more robust model

## Acknowledgements

We thank the Metropolitan Water Reclamation District of Greater Chicago (MWRD), CDPH, CCJ, and ORD for coordination and sample collection.

We also thank Ajit Tamhane (Northwestern University) for helpful conversations.

We thank Eva Durance, Prisila Alvarez, and Haidy Delgado (University of Illinois Chicago) for their contributions to lab work.

All authors were supported by the Illinois Department of Public Health and the Chicago Department of Public Health (CDPH). K.P.L., C.O., M.M.W., A.S., R.P., A.I.P., and N.M.M. were also supported by the Walder Foundation Coronavirus Assessment Network and the University of Illinois system Discovery Partners Institute. M.M.W. was also supported by NSF research training grant DMS-1547394. The conclusions, opinions, or recommendations in this paper are those of the authors and not of IDPH, CDPH, or MWRD.

This project was supported in part by the Centers for Disease Control and Prevention of the U.S. Department of Health and Human Services (HHS) as part of a financial assistance award with 100 percent funded by CDC/HHS. The contents are those of the author(s) and do not necessarily represent the official views of, nor an endorsement, by CDC/HHS, or the U.S. Government.

## References

Adams, C., Bias, M., Welsh, R. M., Webb, J., Reese, H., Delgado, S., Person, J., West, R., Shin, S., & Kirby, A. (2024). The National Wastewater Surveillance System (NWSS): From inception to widespread coverage, 2020–2022, United States. Science of The Total Environment, 924, 171566. https://doi.org/10.1016/j.scitotenv.2024.171566

Berchenko, Y., Manor, Y., Freedman, L. S., Kaliner, E., Grotto, I., Mendelson, E., & Huppert, A. (2017). Estimation of polio infection prevalence from environmental surveillance data. *Science Translational Medicine*, *9*(383), eaaf6786. https://doi.org/10.1126/scitranslmed.aaf6786

Bibby, K., Bivins, A., Wu, Z., & North, D. (2021). Making waves: Plausible lead time for wastewater based epidemiology as an early warning system for COVID-19. *Water Research*, *202*, 117438. https://doi.org/10.1016/j.watres.2021.117438

Centers for Disease Control and Prevention (2023a). National Wastewater Surveillance System: Wastewater Surveillance Data Reporting and Analytics, July 26, 2023. https://www.cdc.gov/nwss/reporting.html

Centers for Disease Control and Prevention (2023b). National Wastewater Surveillance System: Wastewater Surveillance Testing Methods, July 26, 2023. https://www.cdc.gov/nwss/testing.html

Chen, Y., Chen, L., Deng, Q., Zhang, G., Wu, K., Ni, L., Yang, Y., Liu, B., Wang, W., Wei, C., Yang, J., Ye, G., & Cheng, Z. (2020). The presence of SARS-CoV-2 RNA in the feces of COVID-19 patients. *Journal of Medical Virology*, *92*(7), 833–840. https://doi.org/10.1002/jmv.25825

Chik, A. H. S., Glier, M. B., Servos, M., Mangat, C. S., Pang, X.-L., Qiu, Y., D'Aoust, P. M., Burnet, J.-B., Delatolla, R., Dorner, S., Geng, Q., Giesy, J. P., McKay, R. M., Mulvey, M. R., Prystajecky, N., Srikanthan, N., Xie, Y., Conant, B., & Hrudey, S. E. (2021). Comparison of approaches to quantify SARS-COV-2 in wastewater using RT-qPCR: Results and implications from a collaborative inter-laboratory study in Canada. Journal of Environmental Sciences, 107, 218–229. https://doi.org/10.1016/j.jes.2021.01.029

Decaro, N., Elia, G., Campolo, M., Desario, C., Mari, V., Radogna, A., Colaianni, M. L., Cirone, F., Tempesta, M., & Buonavoglia, C. (2008). Detection of bovine coronavirus using a TaqMan-based real-time RT-PCR assay. *Journal of Virological Methods*, *151*(2), 167–171. https://doi.org/10.1016/j.jviromet.2008.05.016

Duan, L., Zhang, Y., Wang, B., Yu, G., Gao, J., Cagnetta, G., Huang, C., & Zhai, N. (2022). Wastewater surveillance for 168 pharmaceuticals and metabolites in a WWTP: Occurrence, temporal variations and feasibility of metabolic biomarkers for intake estimation. *Water Research*, *216*, 118321. https://doi.org/10.1016/j.watres.2022.118321

Feng, S., Owens, S. M., Shrestha, A., Poretsky, R., Hartmann, E. M., & Wells, G. (2023). Intensity of sample processing methods impacts wastewater SARS-CoV-2 whole genome amplicon sequencing outcomes. The Science of the Total Environment, 876, 162572. https://doi.org/10.1016/j.scitotenv.2023.162572

Gushgari, A. J., Venkatesan, A. K., Chen, J., Steele, J. C., & Halden, R. U. (2019). Long-term tracking of opioid consumption in two United States cities using wastewater-based epidemiology approach. *Water Research*, *161*, 171–180. https://doi.org/10.1016/j.watres.2019.06.003

Karthikeyan, S., Nguyen, A., McDonald, D., Zong, Y., Ronquillo, N., Ren, J., Zou, J., Farmer, S., Humphrey, G., Henderson, D., Javidi, T., Messer, K., Anderson, C., Schooley, R., Martin, N. K., & Knight, R. (2021). Rapid, Large-Scale Wastewater Surveillance and Automated Reporting System Enable Early Detection of Nearly 85% of COVID-19 Cases on a University Campus. mSystems, 6(4), 10.1128/msystems.00793-21. https://doi.org/10.1128/msystems.00793-21

Kazama, S., Masago, Y., Tohma, K., Souma, N., Imagawa, T., Suzuki, A., Liu, X., Saito, M., Oshitani, H., & Omura, T. (2016). Temporal dynamics of norovirus determined through monitoring of municipal wastewater by pyrosequencing and virological surveillance of gastroenteritis cases. *Water Research*, *92*, 244–253. https://doi.org/10.1016/j.watres.2015.10.024

Kevill, J. L., Pellett, C., Farkas, K., Brown, M. R., Bassano, I., Denise, H., McDonald, J. E., Malham, S. K., Porter, J., Warren, J., Evens, N. P., Paterson, S., Singer, A. C., & Jones, D. L. (2022). A comparison of precipitation and filtration-based SARS-COV-2 recovery methods and the influence of temperature, turbidity, and surfactant load in urban wastewater. Science of The Total Environment, 808, 151916. https://doi.org/10.1016/j.scitotenv.2021.151916

Leisman, K. P., Owen, C., Warns, M. M., Tiwari, A., Bian, G. (Zhixin), Owens, S. M., Catlett, C., Shrestha, A., Poretsky, R., Packman, A. I., & Mangan, N. M. (2024). A modeling pipeline to relate municipal wastewater surveillance and regional public health data. Water Research, 252, 121178. https://doi.org/10.1016/j.watres.2024.121178

Leonard, B. (2023, October 26). Detecting Covid surges is getting harder, thanks to a contract dispute. POLITICO. https://www.politico.com/news/2023/10/26/detecting-covid-surges-contract-dispute-00123787

Li, W., Su, Y.-Y., Zhi, S.-S., Huang, J., Zhuang, C.-L., Bai, W.-Z., Wan, Y., Meng, X.-R., Zhang, L., Zhou, Y.-B., Luo, Y.-Y., Ge, S.-X., Chen, Y.-K., & Ma, Y. (2020). Virus shedding dynamics in asymptomatic and mildly symptomatic patients infected with SARS-CoV-2. *Clinical Microbiology and Infection: The Official Publication of the European Society of Clinical Microbiology and Infectious Diseases*, *26*(11), 1556.e1-1556.e6. https://doi.org/10.1016/j.cmi.2020.07.008

Lu, X., Wang, L., Sakthivel, S. K., Whitaker, B., Murray, J., Kamili, S., Lynch, B., Malapati, L., Burke, S. A., Harcourt, J., Tamin, A., Thornburg, N. J., Villanueva, J. M., & Lindstrom, S. (2020). *US CDC Real-Time Reverse Transcription PCR Panel for Detection of Severe Acute Respiratory Syndrome Coronavirus 2—Volume 26, Number 8—August 2020—Emerging Infectious Diseases journal—CDC*. https://doi.org/10.3201/eid2608.201246

Metcalf, T. G., Melnick, J. L., & Estes, M. K. (1995). Environmental virology: From detection of virus in sewage and water by isolation to identification by molecular biology--a trip of over 50 years. *Annual Review of Microbiology*, *49*, 461–487. https://doi.org/10.1146/annurev.mi.49.100195.002333

Muttamara, S. (1996). Wastewater characteristics. *Resources, Conservation and Recycling*, *16*(1), 145–159. https://doi.org/10.1016/0921-3449(95)00052-6

Olesen, S. W., Imakaev, M., & Duvallet, C. (2021). Making waves: Defining the lead time of wastewater-based epidemiology for COVID-19. *Water Research*, *202*, 117433. https://doi.org/10.1016/j.watres.2021.117433

Owen, C., Wright-Foulkes, D., Alvarez, P., Delgado, H., Durance, E. C., Wells, G. F., Poretsky, R., & Shrestha, A. (2022). Reduction and discharge of SARS-CoV-2 RNA in Chicago-area water reclamation plants. *FEMS Microbes*, *3*, xtac015. https://doi.org/10.1093/femsmc/xtac015

Kevill, J. L., Pellett, C., Farkas, K., Brown, M. R., Bassano, I., Denise, H., McDonald, J. E., Malham, S. K., Porter, J., Warren, J., Evens, N. P., Paterson, S., Singer, A. C., & Jones, D. L. (2022). A comparison of precipitation and filtration-based SARS-COV-2 recovery methods and the influence of temperature, turbidity, and surfactant load in urban wastewater. Science of The Total Environment, 808, 151916. https://doi.org/10.1016/j.scitotenv.2021.151916

Qian, Q., Fan, L., Liu, W., Li, J., Yue, J., Wang, M., Ke, X., Yin, Y., Chen, Q., & Jiang, C. (2021). Direct Evidence of Active SARS-CoV-2 Replication in the Intestine. *Clinical Infectious Diseases*, *73*(3), 361–366. https://doi.org/10.1093/cid/ciaa925

R Core Team (2021). *R: A language and environment for statistical computing*. R Foundation for Statistical Computing, Vienna, Austria. https://www.R-project.org/

Sheikhzadeh, C. H., Gonzalez-barranca, L., Adams, A. M., Ott, O., Karthikeyan, S., Marotz, L., & Humphrey, G. (2021). *High-throughput Wastewater SARS-CoV-2 Detection Pipeline*. https://www.protocols.io/view/high-throughput-wastewater-sars-cov-2-detection-pi-bshvnb66

Steele, J. A., Zimmer-Faust, A. G., Griffith, J. F., & Weisberg, S. B. (2021). *Sources of variability in methods for processing, storing, and concentrating SARS-CoV-2 in influent from urban wastewater treatment plants* (p. 2021.06.16.21259063). medRxiv. https://doi.org/10.1101/2021.06.16.21259063

v. Wintzingerode, F., Göbel, U. B., & Stackebrandt, E. (1997). Determination of microbial diversity in environmental samples: Pitfalls of PCR-based rRNA analysis. FEMS Microbiology Reviews, 21(3), 213–229. https://doi.org/10.1016/S0168-6445(97)00057-0

Wade, M. J., Lo Jacomo, A., Armenise, E., Brown, M. R., Bunce, J. T., Cameron, G. J., Fang, Z., Farkas, K., Gilpin, D. F., Graham, D. W., Grimsley, J. M. S., Hart, A., Hoffmann, T., Jackson, K. J., Jones, D. L., Lilley, C. J., McGrath, J. W., McKinley, J. M., McSparron, C., … Kasprzyk-Hordern, B. (2022). Understanding and managing uncertainty and variability for wastewater monitoring beyond the pandemic: Lessons learned from the United Kingdom national COVID-19 surveillance programmes. *Journal of Hazardous Materials*, *424*, 127456. https://doi.org/10.1016/j.jhazmat.2021.127456

Warwick, C., Guerreiro, A., & Soares, A. (2013). Sensing and analysis of soluble phosphates in environmental samples: A review. *Biosensors and Bioelectronics*, *41*, 1–11. https://doi.org/10.1016/j.bios.2012.07.012

Wolfe, M. K., Duong, D., Bakker, K. M., Ammerman, M., Mortenson, L., Hughes, B., Arts, P., Lauring, A. S., Fitzsimmons, W. J., Bendall, E., Hwang, C. E., Martin, E. T., White, B. J., Boehm, A. B., & Wigginton, K. R. (2022). Wastewater-Based Detection of Two Influenza Outbreaks. *Environmental Science & Technology Letters*, *9*(8), 687–692. https://doi.org/10.1021/acs.estlett.2c00350

Wolfe, M. K., Yu, A. T., Duong, D., Rane, M. S., Hughes, B., Chan-Herur, V., Donnelly, M., Chai, S., White, B. J., Vugia, D. J., & Boehm, A. B. (2023). Use of Wastewater for Mpox Outbreak Surveillance in California. *New England Journal of Medicine*, *388*(6), 570–572. https://doi.org/10.1056/NEJMc2213882

Zheng, X., Deng, Y., Xu, X., Li, S., Zhang, Y., Ding, J., On, H. Y., Lai, J. C. C., In Yau, C., Chin, A. W. H., Poon, L. L. M., Tun, H. M., & Zhang, T. (2022). Comparison of virus concentration methods and RNA extraction methods for SARS-COV-2 wastewater surveillance. Science of The Total Environment, 824, 153687. https://doi.org/10.1016/j.scitotenv.2022.15368